\documentclass[preprint]{aastex}

\newcommand{\h}{H\scriptsize II \normalsize}
\newcommand{\etal}{et al.}
\newcommand{\3}{[O III]~$\lambda$5007}
\newcommand{\1}{[O I]~$\lambda$6300}

\begin{document}            

\title{THE PRIMORDIAL HELIUM ABUNDANCE: TOWARDS UNDERSTANDING AND
REMOVING THE COSMIC SCATTER IN THE $dY/dZ$ RELATION}

\author{D. R. Ballantyne\altaffilmark{1}, G. J. Ferland\altaffilmark{2} 
and P. G. Martin}
\affil{Canadian Institute for Theoretical Astrophysics, University of
Toronto, Toronto, ON, Canada~M5S~3H8; ballanty, ferland and
pgmartin@cita.utoronto.ca}

\altaffiltext{1}{Current address: Institute of Astronomy, 
Madingley Road, Cambridge, United Kingdom CB3 0HA}

\altaffiltext{2}{Also: Department of Physics and Astronomy, 
University of Kentucky, Lexington, KY 40506-0055 USA} 

\begin{abstract}
We present results from photoionization models of low-metallicity \h
regions. These nebulae form the basis for measuring the primordial helium abundance.
Our models show that the helium ionization correction factor (ICF) can be non-negligible
for nebulae excited by stars with effective temperatures larger than 40,000~K. 
Furthermore, we find that when the effective temperature rises to above 45,000~K, the
ICF can be significantly negative. This result is independent of the choice of stellar
atmosphere. However, if an \h region has an \3/\1 ratio greater than 300, then our 
models show that, regardless of its metallicity, it will have a negligibly small ICF. 
A similar, but metallicity dependent, result was found using the \3/H$\beta$ ratio. These
two results can be used as selection criteria to remove nebulae with potentially 
non-negligible ICFs. Using our metallicity independent criterion on the data of
\citet{izo98} results in a 20\% reduction of the rms scatter about the best fit 
$Y-Z$ line. A fit to the selected data results in a slight increase of the value 
of the primordial helium abundance.  

\end{abstract}


\section{INTRODUCTION}
\label{sec:intro}

An accurate measurement of the primordial helium abundance would be
an important test of standard big bang nucleosynthesis \citep{oli97}, and
would also constrain the values of the photon-to-baryon ratio and $\Omega_b$
 \citep{oli99}. The traditional procedure to measure the primordial helium
abundance is to make use of the correlation between the helium mass fraction
 ($Y$) and metal abundance ($Z$). This correlation is then extrapolated
to zero metallicity to estimate the primordial mass fraction of helium, 
$Y_p$. Spectroscopic observations of bright, low-metallicity extragalactic 
\h regions provide the data for these studies \cite[e.g.,][]{oli95,oli97,
izo94,izo97,izo98,tor89,skil98}. 

To be cosmologically useful the value of $Y_p$ has to be determined to better
than 5\%. Fortunately, abundance determinations from measurements of line ratios
 is relatively straightforward \citep{pei75,ben99}, and can, {\it in theory}, 
give the desired accuracy. However, to reach the needed level of precision, any
systematic errors involved with target selection, observations, and data 
analysis must be identified and corrected. Many such systematic 
errors have already been identified
\citep{dav85,din86,pag92,skil94,pei96,izo97,ste97,skil98}, but any errors resulting 
from the so-called ``ionization correction factor'' (ICF) have so far been 
assumed to be small. The ICF corrects for the fact that some amount of atomic 
(i.e., unseen) helium might be present in ionized regions of hydrogen 
\citep{ost89,pei75}. This correction traditionally has been assumed to be 
zero because measurements of the primordial helium abundance employ 
observations of bright extragalactic \h regions. These regions are excited by 
clusters of young stars with effective temperatures greater than
40,000~K. Calculations by \citet{sta90} and \citet{pag92} showed that the
helium ICF should be negligibly small for these \h regions. As a result, 
recent determinations of $Y_p$ have assumed that the helium ICF is small.

Very recently, \citet{arm99} presented calculations that showed that
\h regions excited by stars with temperatures greater than 40,000~K can
have non-negligible ICFs. \citet{arm99} found that the ICFs were often negative 
(i.e., the helium ionized zone is {\bf larger} than the hydrogen one; 
\citet{sta80,sta82}, \citet{pen86}) for the hardest 
stellar continua. These results were confirmed by \citet{vie99}. 
In this paper, we follow up on the work of \citet{arm99}, and develop observational 
diagnostics of when the He ICF is important and when it can be ignored. We then 
apply these diagnostics to the data of \cite{izo98} to illustrate how our technique
 can improve the precision of the measurement of $Y_p$. 

We describe our calculations in \S~\ref{sec:calc}, and our results in
\S~\ref{sec:results}. The main results are summarized in \S~\ref{sec:concl}.

\section{DESCRIPTION OF CALCULATIONS}
\label{sec:calc}

In order to investigate the effects of a non-negligible ICF on the determination of the 
primordial helium abundance, we ran photoionization models of \h regions and extracted 
the ICF for each nebula. These calculations are very similar to ones presented by 
\citet{arm99} and \citet{bot98}, and were made with the development version 
of Cloudy, last described by \citet{fer98}.

Since we are modeling \h regions, our models use the ISM abundances and grain model that 
were used and described by \citet{arm99}. However, we scaled both the metal 
and grain abundances to lower values because, in this case, we are most interested in 
lower metallicity nebulae. The scaling was implemented so that all metals and grains
were varied together relative to hydrogen and helium, but the He/H ratio was held
constant. We modeled nebulae at three different metallicities: O/H=32, 64, and 128 
parts per million [ppm] ($Z=Z_{\odot}/23, Z_{\odot}/12$, and $Z_{\odot}/6$ respectively). 
For each metallicity, 1936 models were computed for each
of the following spectral energy distributions: the LTE plane-parallel
atmospheres of \citet{kur91}, the non-LTE, wind-blanketed, solar
abundance CoStar atmospheres of \citet{sc96a,sc96b}, the earlier NLTE
atmospheres of \citet{mih72}, and for completeness blackbodies.  We also
ran models using the subsolar-abundance CoStar atmospheres which are
spectrally slightly harder than the solar abundance ones. These models
resulted in slightly more negative ICFs, but the values of the line
ratio cutoffs (\S~\ref{sub:zdep} \& \S~\ref{sub:zindep}) were not
changed from the ones calculated with a solar abundance atmosphere.

For each atmosphere we computed models with 10~cm$^{-3} \leq n_H \leq 10^6$~cm$^{-3}$, 
10$^{-4} \leq U \leq 10^{-0.25}$, and 40,000~K$ \leq T_{eff} \leq 50,000$~K, where
$U$ is the ionization parameter defined as in Eq.~4 of \citet{arm99}. Giant 
extragalactic \h regions that are observed are generally excited by large clusters, so this 
range of parameters should cover all such nebulae. We modeled the nebulae as
plane-parallel constant density slabs, a simple way to characterize blister \h regions.
Our proposed diagnostic indicators are the \3 and \1 lines (\S~\ref{sub:zdep} \& 
\S~\ref{sub:zindep}), and these should be fairly independent of the
assumed geometry (sphere, sheet, or evaporating blister). The \3/H$\beta$
ratio represents the cooling per recombination (the Stoy ratio) and so is
primarily sensitive to the stellar temperature \citep{sto33,kal78}
rather than geometry.  Similarly, the \1/H$\beta$ intensity ratio mostly
measures the "softness" of the hydrogen ionization front (where the line
forms - Netzer \& Davidson 1979).

\section{RESULTS}
\label{sec:results}
\subsection{Temperature Dependence}
\label{sub:temp}

Figure~\ref{fig:temp-icf} plots the ICF calculated from the Kurucz and CoStar models 
versus the stellar temperature. Note that we define the ICF such that an ICF of zero 
corresponds to zero correction:
\begin{equation}
\label{eq:ICF}
\mathrm{ICF}={\left < \mathrm{H^{+}}\!/\,\mathrm{H} \right > \over \left < 
\mathrm{He^{+}}\!/\,\mathrm{He} \right > } - 1,
\label{eq:icf}
\end{equation}
where the angle brackets denote the volume mean ionization fraction. This definition
takes into account the presence of any He$^{+2}$ in the nebula. 
Figure~\ref{fig:temp-icf} clearly 
shows that one can obtain a non-negligible ICF for stars with temperatures greater 
than 40,000~K. The harder CoStar atmospheres give preferentially negative ICFs, which, 
if not taken into account, would result in a overestimate of the helium abundance. The 
same is true of the Mihalas atmospheres (not shown). These results agree with the 
calculations of \citet{arm99} and \citet{vie99}. The softer Kurucz atmospheres result in 
preferentially positive ICFs (i.e., the helium ionized zone is smaller than the hydrogen one), 
although, at temperatures greater than 45,000~K, they can 
also give negative ICFs. The blackbody atmospheres, the least realistic, are softer still, 
but even they can produce negative ICFs in some models at the highest temperatures. 
Therefore, negative ICFs seem to be found at high stellar temperatures {\it independent 
of the type of stellar atmosphere}.

\subsection{Metallicity Dependent Cutoff Criterion}
\label{sub:zdep}

A negative ICF occurs as the results of penetrating high-energy photons 
preferentially ionizing helium, due to its large photoionization cross section. 
This tends to be important for lower ionization parameter models, since these have
significant regions where H and He are partially ionized. We expect these nebula to be
characterized by lower \3/H$\beta$ ratios (a measure of excitation) and larger
\1/H$\beta$ ratios (since \1 is formed in warm atomic regions).

The results presented in \S~\ref{sub:temp} show that it is not appropriate to simply 
assume that the ICF is zero when a nebula is excited by a star with a temperature 
greater than 40,000~K. However, it would be important to develop observational 
diagnostics for when the ICF is important and when it is not. Figure~\ref{fig:o3-icf} 
shows such a diagnostic. A plot of ICF versus \3/H$\beta$ shows that beyond a line 
ratio of about 3--4 the ICF is negligible (for clarity results are shown for Kurucz 
and CoStar atmospheres only; a plot for blackbody and Mihalas atmospheres is very
similar). However, the value of the cutoff will depend on metallicity. We found very 
small ICFs for line ratios greater than the following cutoff:
\begin{equation}
\label{eq:zcutoff}
\left (\mathrm{[O\ III]}\ \lambda 5007/\mathrm{H}\beta \right )_{\mathrm{Cutoff}} =
(0.025 \pm 0.004)\mathrm{O/H}+(1.139 \pm 0.306),
\end{equation}
where O/H is measured in parts per million. \h regions which have an \3/H$\beta$
ratio less than the cutoff for their metallicity might be subject to an ICF 
correction. Unless this correction can be made (and, in general, it cannot) these
\h regions should be removed from the abundance analysis, as they will increase the
scatter in the $dY/dZ$ relation used to determine the primordial helium abundance.

\subsection{Metallicity Independent Cutoff Criterion}
\label{sub:zindep}

One can improve the above result by finding emission line ratios that should be
independent of metallicity. Figure~\ref{fig:o3o1-icf} plots the helium ICF versus
the \3/\1 ratio. The figure shows that the ICF is negligible for a line ratio greater
than about 300. Not surprisingly, this result is {\it independent of metallicity}. 

There are some Kurucz and blackbody models which result in a non-negligible ICF at line 
ratios larger than this cutoff. These models had low stellar temperatures 
(40,000--42,000~K), and were found over a narrow range in both $\log U$ ($-1.5$ to $-2.25$) 
and $\log n_H$ (1.0--3.0). Their positive ICF is a result of combining the low stellar
temperatures and the softness of their atmospheres; their large \3/\1 ratio is a result
of combining the fairly high ionization parameter with the low density. CoStar and Mihalas 
models, which are considered more ``realistic'', result in only very small ICFs in this 
region.

Therefore, we find that any \h region, regardless of its metallicity, that has an
\3/\1 ratio less than about 300 might be subject to an ICF correction and should not 
be used to determine the primordial helium abundance.

\subsection{Application to Real Data}
\label{sub:apply}

To see how these new results affect the determination of $Y_p$, we applied the
metallicity independent rejection criteria to the data of \citet{izo98}, and the
results are shown in Figure~\ref{fig:o3o1-y}. There are a number of points to note
from this Figure:
\begin{enumerate}

\item Our criterion rejects points over the entire range of metallicity, so there is
no metallicity bias. The rejected points also fall evenly over the range of $Y$ values,
which implies that there is no correlation between $Y$ and the \3/\1 ratio.

\item At an \3/\1 cutoff of 300 there is a 20\% reduction in the weighted rms scatter 
about the best fit line. This is encouraging evidence that part of the scatter was
due to the ICF, and the situation has indeed improved by the implementation of the
cutoff.

\item The negative slopes predicted by using a
cutoff$\ \geq$ 300 result from weighted fits to the small number of data
that remain after applying the cutoff, and are probably not
realistic. Given the current data, a cutoff value greater than 300 is
too severe (not practical).

\item Despite the increased errors in the slope and the intercept, we find that
implementing our cutoff will result in a larger value of $Y_p$. Specifically, we
find $Y_p = 0.2489 \pm 0.0030$ when the cutoff is taken at a \3/\1 ratio of 300. 
\cite{izo98} find $0.2443 \pm 0.0015$ (our fit gives $0.2443 \pm 0.0013$ with no cutoff), 
so the two results are barely consistent at the 1$\sigma$ level. This result is 
in the opposite direction of the shift predicted by the Monte Carlo simulations of 
\citet{vie99}, but moves the value of $Y_p$ closer to the theoretically predicted 
values \citep{oli95}.

\end{enumerate}

There is always the possibility that systematic errors are introduced whenever data
are rejected by a certain criterion. The above selection criterion preferentially
selects \h regions that have large \3/\1 ratios. This ratio is generally large 
whenever the \3 line is strong (this is consistent with our metallicity dependent 
criterion). The \3 line is a major source of cooling in a nebula, and so by selecting 
\h regions with strong \3 lines we are selecting regions ionized by hotter stars. Since 
these large extragalactic \h regions are generally ionized by clusters, ones with 
strong \3 lines are preferentially younger. However, this is unlikely to introduce a 
systematic error in the primordial helium abundance determination, as young clusters 
can form at any metallicity. Indeed, Figure~\ref{fig:o3o1-y} shows that the points 
rejected by applying the cutoff span the entire range of metallicity.

There is also the possibility that physical conditions within the \h regions
may bias our results. For example, in some nebulae the intensity of the \1 line 
could be enhanced due to shock heating. This would lower the measured \3/\1 ratio,
and could move it below our selection criterion. However, shock heating would not
change the ICF of the nebula, so even though application of our criterion might
reject such a \h region, it will not bias the determination of $Y_p$. 

Another potential situation in \h regions is that the nebula may be matter
bounded (i.e., optically thin to the Lyman continuum) in certain solid
angles or sectors.  Because there will be no hydrogen ionization front in such
sectors, and therefore no \1 line emission, the presence of such sectors
will increase the measured \3/\1 ratio, possibly pushing it above our selection 
criterion. However, these sectors will also have little or no neutral helium 
within them, and so will have no ICF. Therefore, although the
other, ionization bounded, sectors of the \h region could have a
non-negligible ICF, this will be diluted by the matter bounded sectors.
We anticipate that even if such an \h region were shifted into our
selected data, there ought not be a large effect on determining $Y_p$.  According
to Fig.~\ref{fig:o3o1-icf}, to severely bias the results a number of points would have 
to shift rightward in \3/\1 by a factor larger than ten; but such a large shift in 
the line ratio would probably result in a large dilution in the ICF.  More modeling 
would be needed to quantify how little an impact matter bounded sectors would have on 
the \3/\1 ratio and ICFs.

\section{CONCLUSIONS}
\label{sec:concl}
In this paper we have shown the following:
\begin{enumerate}

\item There can be a non-negligible ICF correction for \h regions excited by stars with
temperatures greater than 40,000~K. At temperatures higher than 45,000~K, the ICF is
preferentially negative. This result is independent of the atmosphere of the O star.

\item There is a simple procedure to determine if an ICF correction needs to be made for
a given \h region. If the \3/\1 ratio is greater than 300, then no correction is needed.
This criterion is independent of metallicity. If the \1 line cannot be measured, then 
there is a metallicity dependent cutoff (Eq.~\ref{eq:zcutoff}) that can be used 
with the \3 line.

\item Applying the metallicity independent criterion to the data of \citet{izo98}
results in reducing the rms scatter about the best fit $Y-Z$ line by 20\%. 
This will help remove systematic errors relating to unrecognized ICF effects, and ought to 
improve the reliability of the $Y_p$ determination. Furthermore, an analysis of the 
selected data gives a larger value of $Y_p$ than was originally measured, which is 
closer to the theoretically expected value.
\end{enumerate}   

\acknowledgements

G.J.F.\ thanks CITA for its hospitality during a sabbatical year and acknowledges
support from the Natural Science and Engineering Research Council of Canada through
CITA. D.R.B.\ also acknowledges financial support from NSERC. We acknowledge useful
comments from an anonymous referee.

\clearpage


\figcaption{A plot of the He ICF values obtained from the CoStar and Kurucz 
photoionization models versus stellar temperature of the atmospheres. Only points 
with ICF between $\pm$10\% are plotted. There are non-negligible ICFs at all values
of stellar temperature. The CoStar atmospheres generally result in negative ICFs, 
while the softer Kurucz atmospheres give positive ICFs for temperatures less than about
45,000~K. These models were run with a metallicity of (O/H)=64 ppm. \label{fig:temp-icf}}

\figcaption{A plot of He ICF vs. \3/H$\beta$ using data from the CoStar and 
Kurucz grids run at (O/H)=64 ppm. Only points with ICF between $\pm$10\% were 
plotted. At this metallicity the ICF-related cutoff is at a relative line strength of 
about 3--4. See the text for discussion on how cutoff varies with metallicity. 
\label{fig:o3-icf}} 

\figcaption{ a) A plot of He ICF vs. \3/\1 with data obtained from the
CoStar and Kurucz model grids run at (O/H)=64 ppm. The ideal ICF-related
cutoff here is at a line ratio of 300. This cutoff is {\it independent}
of metallicity. The Kurucz models that have positive ICFs in the allowed
zone are models with a particular combination of parameters (see text). 
b) Like a) but for Mihalas and blackbody atmospheres.
\label{fig:o3o1-icf}}

\figcaption{This figure shows the effects of applying the metallicity independent
rejection criterion to the data of Izotov \& Thuan (1998). The solid line is a 
weighted least-squares fit to the selected data points shown by the solid symbols. 
The open symbols are the rejected points. For reference, the dashed line is the fit 
with no cutoff applied. Note that as the \3/\1 cutoff becomes larger, the scatter of 
the points about the best-fit line becomes smaller (there is a 20\% reduction in
the rms scatter when the cutoff is 300). This procedure also shows that the value of 
$Y_p$ determined by Izotov \& Thuan (1998) might be an underestimate. 
\label{fig:o3o1-y}}

\clearpage

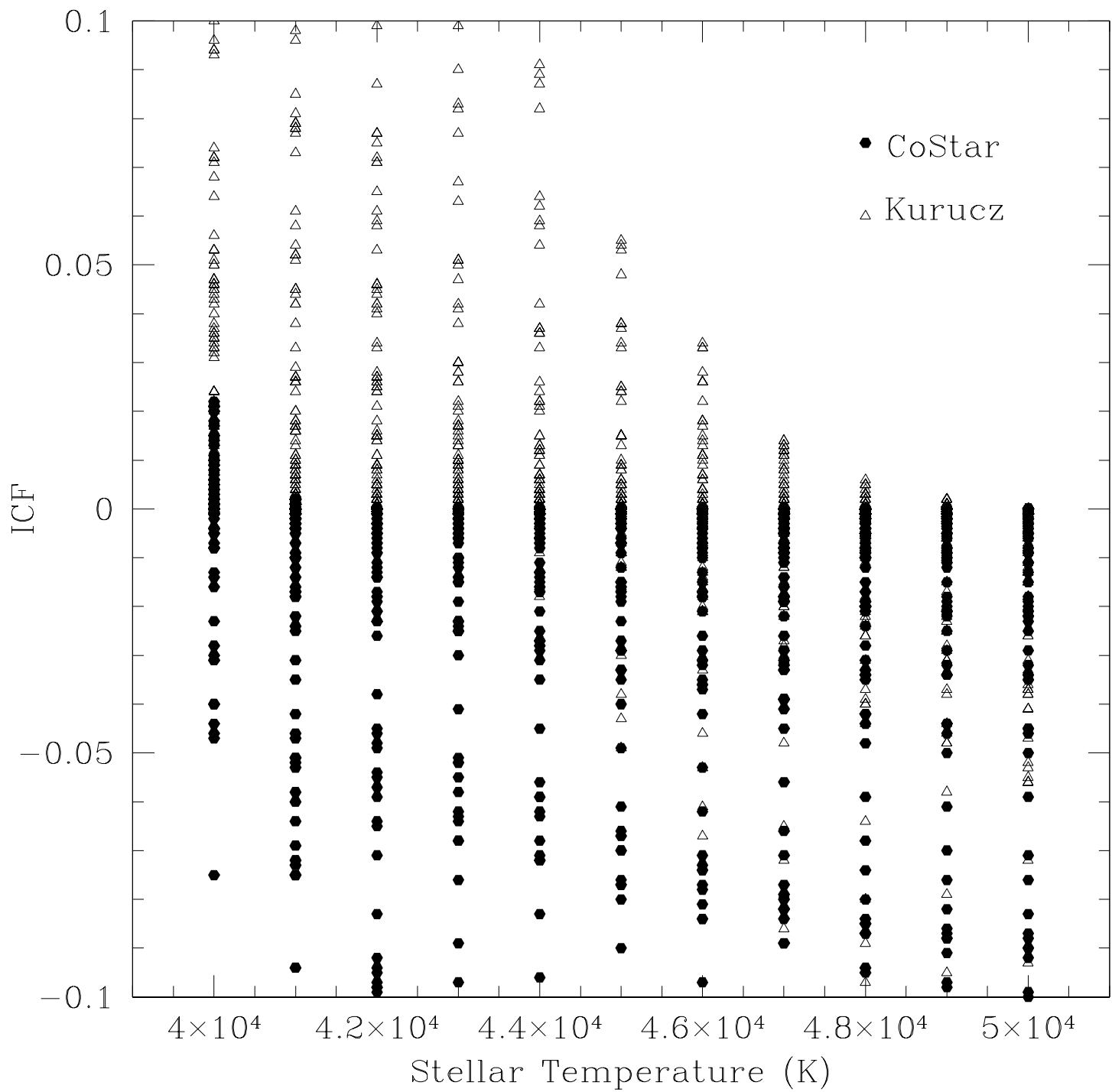
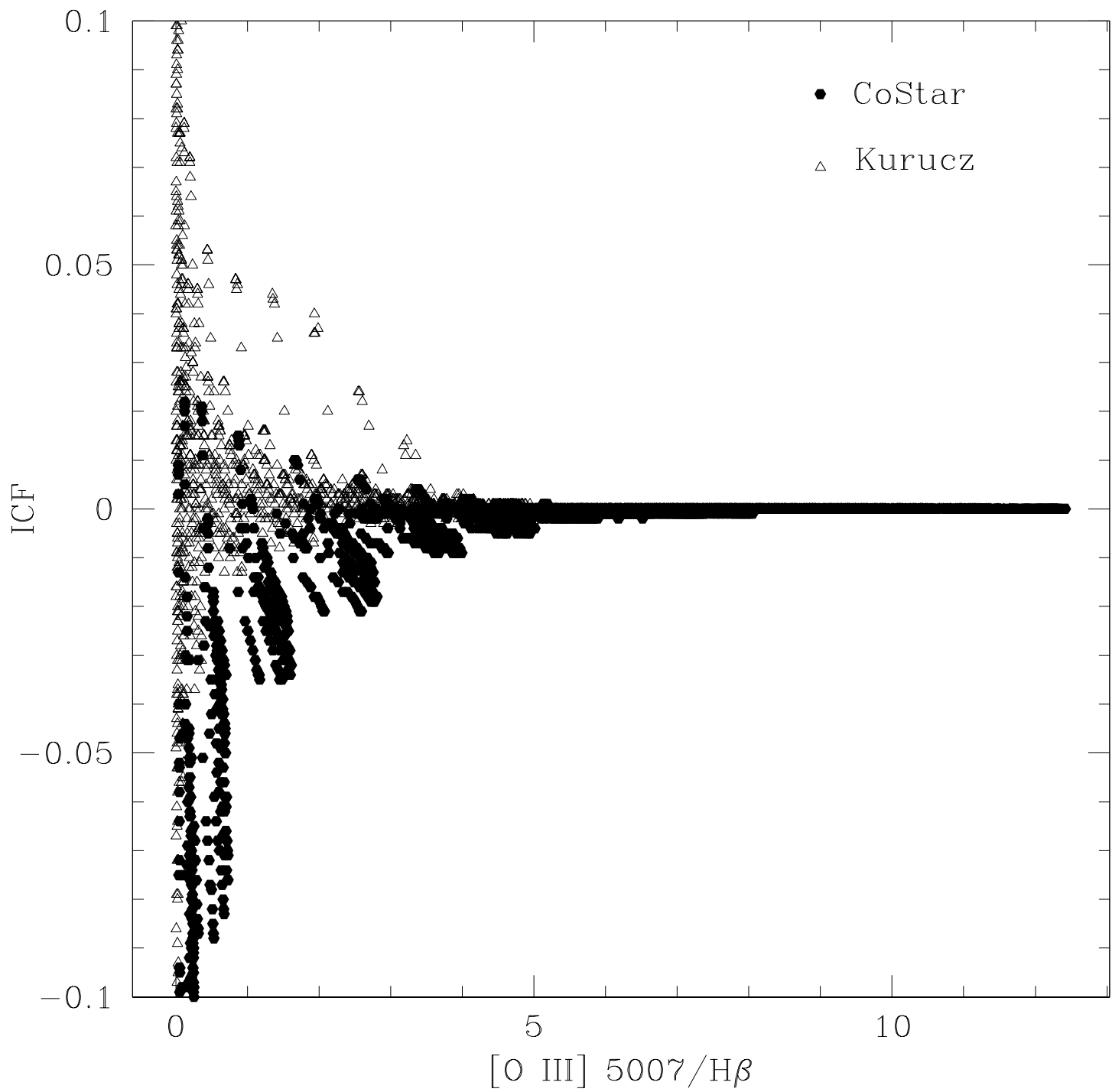
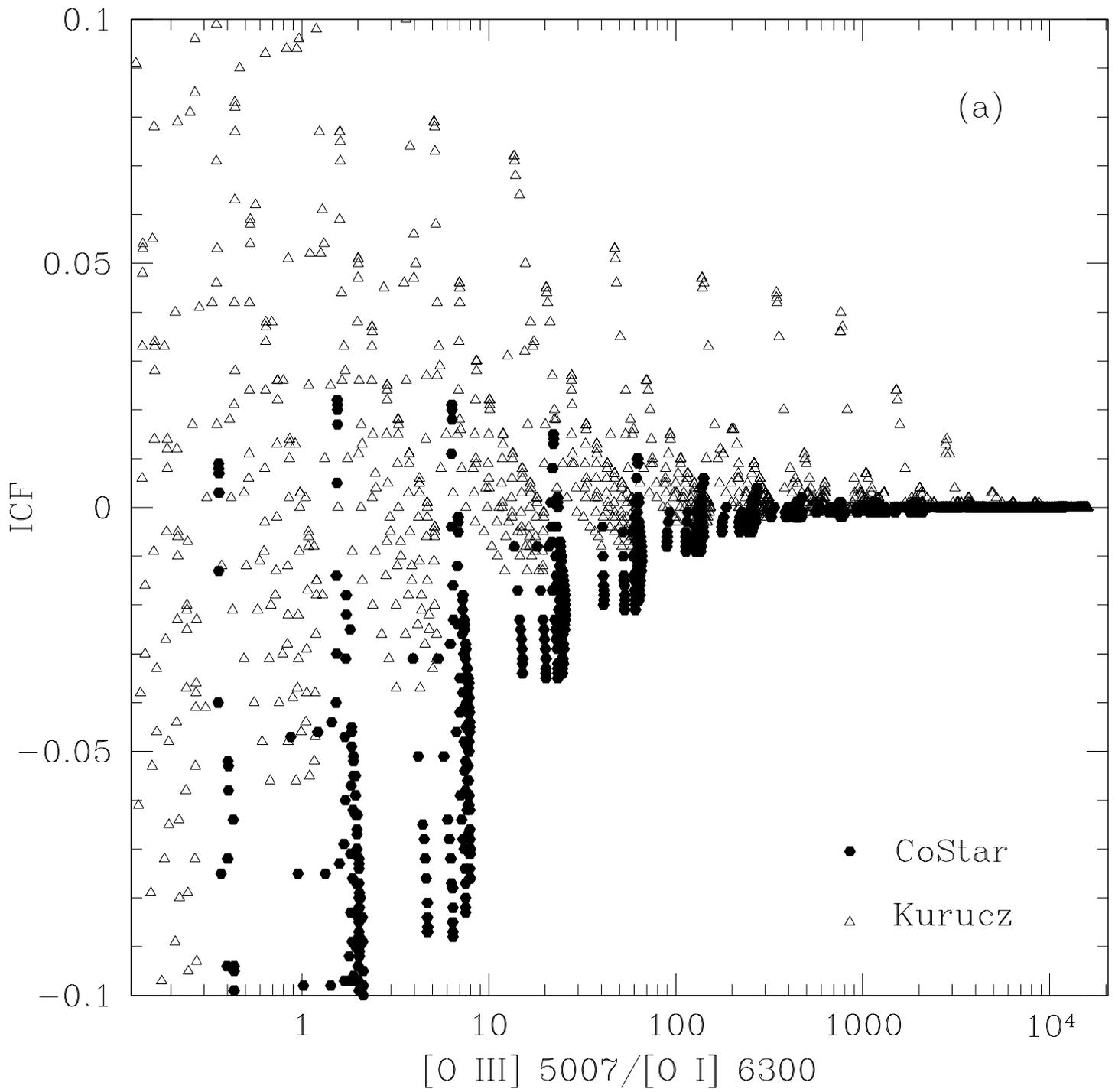
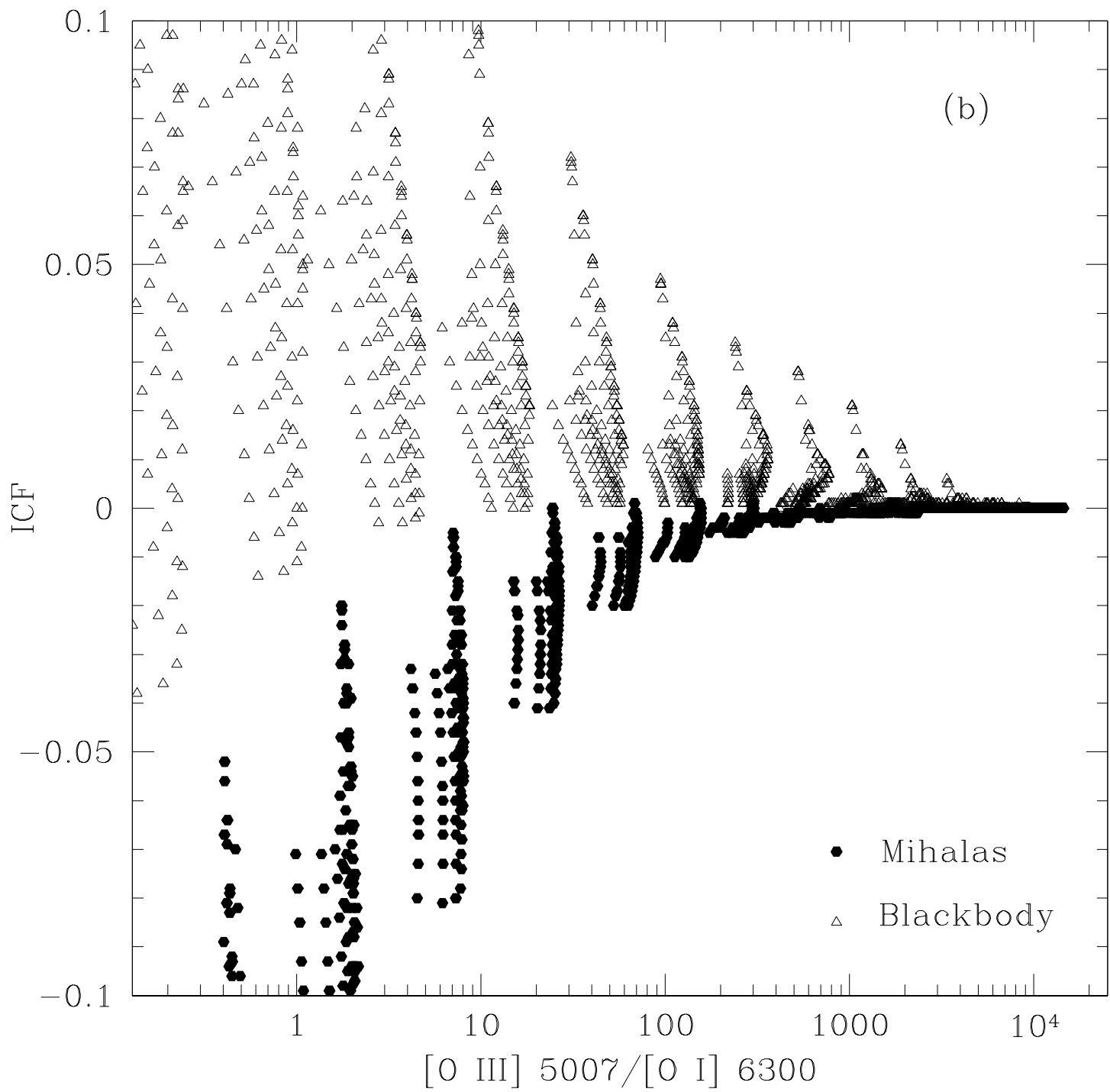
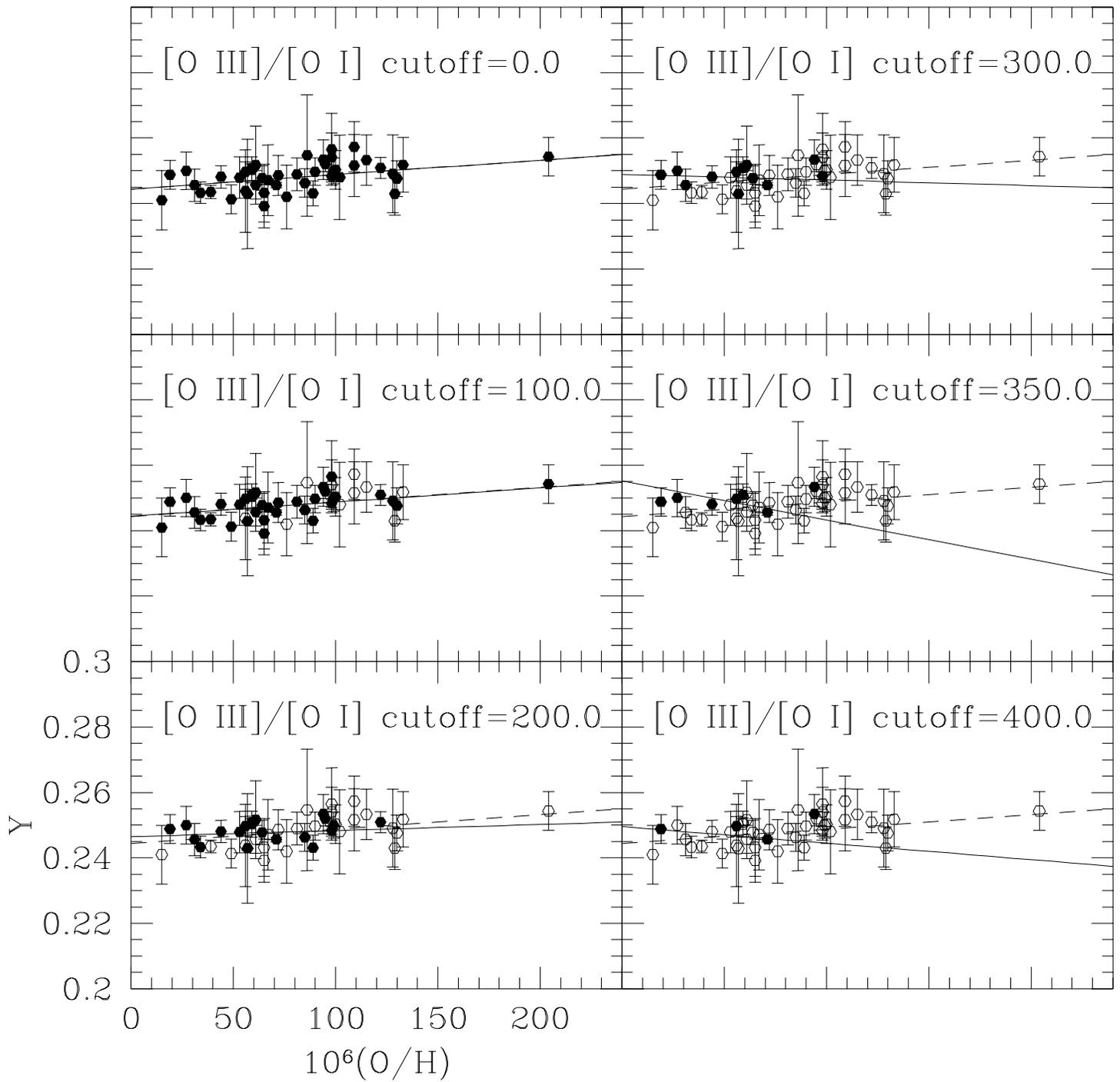
\end{document}